\documentclass{ws-mpla}

\newcommand{\be}{\begin{equation}}
\newcommand{\ee}{\end{equation}}
\newcommand{\bea}{\begin{eqnarray}}
\newcommand{\eea}{\end{eqnarray}}

\begin{document}

\markboth{Gero von Gersdorff}
{Flavor Physics in Warped Geometry}
\catchline{}{}{}{}{}

\title{Flavor Physics in Warped Space}

\author{\footnotesize GERO VON GERSDORFF}

\address{
ICTP South American Institute for Fundamental Research
Instituto de Fisica Teorica,\\
 Sao Paulo State University,
S\~ao Paulo, SP,
Brazil\\
gersdorff@gmail.com}



\maketitle

\pub{Received (Day Month Year)}{Revised (Day Month Year)}

\begin{abstract}
We review constraints from quark and lepton flavor violation on extra dimensional models with warped geometry, both in the minimal and the custodial model. For both scenarios, KK masses that are large enough to suppress constraints from electroweak precision tests also sufficiently suppress all quark flavor and CP violation, with the exception of CP violation in $K\bar K$ mixing and (to a  lesser extend) in $D\bar D$ mixing. In the lepton sector the minimal scenario leads to excessively large contributions to $\mu\to e\gamma$ transitions, requiring KK masses of at least 20 TeV or larger.
\end{abstract}


\section{Introduction}

Models of warped extra dimensions\cite{Randall:1999ee} (WED) have become one of the major ideas in addressing the naturalness problems of the Standard Model (SM), and have been the subject of many phenomenological studies over the past years. As it is the case with any new physics appearing at the low scale, there exists a potential flavor problem.
However, it was soon realized that localizing matter in the bulk of the extra dimension\cite{Grossman:1999ra,Gherghetta:2000qt,Huber:2000ie} not only opens the possibility to a natural explanation of the SM fermion mass hierarchy, but also offers a very effective suppression of flavor violating effects.

This brief review is meant as a summary of the status of flavor violation in WED and the leading bounds coming from both quark and lepton transitions. In this work we will not consider at all models with all fields residing on the IR brane. Such models are in principle much more severely constrained, as quantum gravity effects induce (incalculable) IR brane localized flavor violation at the TeV scale.
The paper is structured as follows. In Sec.~\ref{warpedflavor} we briefly recapitulate the flavor structure generated by WED with matter in the bulk, and review the main constraints coming from electroweak precision tests (EWPT).
In section \ref{operators} we identify the potentially dangerous flavor violating operators as they are generated at the scale of the Kaluza Klein (KK) masses. Sec.~\ref{boundsquarks} focuses on $\Delta F=2$ and $\Delta F=1$ transitions in the quark sector, and in Sec.~\ref{boundsleptons} we consider lepton flavor violation (LFV).


\section{Warped Theories of Flavor}
\label{warpedflavor}

Warped theories of flavor are based on the localization of fermion fields  at different points in the extra dimension.\cite
{Grossman:1999ra,Gherghetta:2000qt,Huber:2000ie} The model is defined by metric \cite{Randall:1999ee}
\be
ds^2=\frac{1}{(kz)^2}(dx_\mu^2+dz^2)
\ee
with ultraviolet (UV) and infrared (IR) branes located at $z_0=k^{-1}$ and $z_1=\tilde k^{-1}$ respectively. The quantity $\epsilon\equiv \tilde k/k$ is known as the warp factor and should be fixed to $\epsilon\sim 10^{-16}$ in order to account for the full Planck-TeV hierarchy. The masses of the lightest gauge-KK resonances (including possible custodial partners) are given by $m_{\rm KK}\approx 2.4\, \tilde k$. The fermion equation of motion determines their normalized zero mode profiles $f_f(z)$
\be
\Psi_f(x,z)=f_f(z)\, \psi_f(x)+{\rm KK\ excitations}
\ee
in terms of their bulk mass\footnote{We take  fields with left handed zero modes ($f=q_L,\ell_L$) to  have  bulk mass  $m_{f}=- c_f\, k$ and those with right handed zero modes ($f=u_R,d_R,e_R$) to have bulk mass  $m_{f}=+ c_f\, k$. In this convention, all fields with $c_f>\frac{1}{2}$ are UV localized.
} $m_{f}=\mp c_f\, k$ as
\be
f_f(z)=\sqrt{(1-2c_f)k}\frac{(kz)^{2-c_f}}{\sqrt{(kz_1)^{1-2c_f}-1}}
\ee
The profiles of the Higgs field zero mode is given by
\be
f_\phi(z)=\sqrt{2(a-1)k}\frac{(kz)^a}{\sqrt{(kz_1)^{2(a-1)}-1}}\,\qquad a>2
\ee
where the inequality ensures that the Higgs field is sufficiently IR localized and thus the hierarchy problem is solved by the warping. The parameter $a$ determines the localization of the Higgs field, in particular $a=2$ corresponds to gauge-Higgs unification and $a=\infty$ to a brane Higgs.  \footnote{In case of an  5d bulk Higgs field, $a$ can be related to its bulk mass as $a=2+\sqrt{4+m_\phi^2/k^2}$.} Yukawa couplings are then computed using wave function overlap integrals. In practice, the only relevant regime is when $a>c_{q_L}+c_{q_R}$, in which case the Yukawas can be approximated as
\be
Y^q_{ij}\sim 
\hat Y^q_{ij} \epsilon_{q_L^i}\epsilon_{q_R^j} \,,\qquad \epsilon_f=\sqrt{\frac{1-2c_f}{1-\epsilon^{1-2c_f}}}
\ee
where $\hat Y$ denote the 5d Yukawa couplings in units of the curvature $k$.
Fermion Yukawa hierarchies can then arise purely from $\mathcal O(1)$ numbers by localizing all but the third generation quark doublet and the right handed (RH) top near the UV brane, $c_f>\frac{1}{2}$. Typical values obtained from the fit in Ref.~\refcite{Cabrer:2011qb} that reproduce the known quark masses and mixings are given in Tab.~\ref{ta1}.  
\begin{table}[h]
\tbl{ Medians and $1\sigma$ confidence intervals of the $c$ parameters corresponding to the different species of quarks and chiralities, with $a=2$.}
{\begin{tabular}{@{}lll@{}} \toprule
$c_{q^1_L} =0.66 \pm 0.02$ & $c_{q^2_L} = 0.59 \pm 0.02  $ & $c_{q^3_L} =  -0.11^{+0.45}_{-0.53} $
\\  
$c_{u_R} = 0.71 \pm 0.02$ & $c_{c_R} = 0.57 \pm 0.02$ & $c_{t_R} =  0.42^{+0.05}_{-0.17}$
\\ 
$c_{d_R} = 0.66 \pm 0.03$ & $c_{s_R} = 0.65 \pm 0.03 $ & $c_{b_R} = 0.64 \pm 0.02$ 
\\ 
\botrule
\end{tabular}\label{ta1} }
\end{table}

We would like to point out two fine tuning issues with the choice of the $c$ parameters, which are not always very much appreciated in the literature.
\begin{itemize}
\item All the UV localized fields (with $c_f>\frac{1}{2})$ need to have values $c_f$ rather close to the critical value $\frac{1}{2}$. The reason for this is that the suppression factors behave as $\epsilon_{f}\sim\epsilon^{c_f-\frac{1}{2}}$, i.e.~they scale as powers of the Planck-TeV hierarchy $\epsilon=\tilde k/k$, which is rather large compared to the typical flavor hierarchies.
\item
The $c_{d_R^i}$ are quite degenerate. This lack of right handed hierarchy results typically in large right handed down quark rotations, making flavor observables involving these fields particularly sensitive to the KK mass scale.\footnote{The RS-GIM mechanism explained below implies that flavor violating couplings scale as $\Delta c\, \epsilon^{|\Delta c|}$, and hence they are maximal for $\Delta c=\left|\log \epsilon\right|^{-1}\approx 0.027$.}
\end{itemize}
The reason for the second point is that the hierarchy in the $\epsilon_{q_L^i}$ is completely determined by the CKM matrix as\footnote{A useful relation to remember is that all rotations roughly scale as $V^q_{ij}= \frac{\epsilon_{q^j}}{\epsilon_{q^i}}$, for $i<j$ .}
\be
\epsilon_{q_L^1}:\epsilon_{q_L^2}:\epsilon_{q_L^3}\sim 1:5:125.
\ee
Using this in the hierarchy for the eigenvalues of the up and down quark Yukawa couplings,
$y_d:y_s:y_b\sim 1:20:800$
\ and $y_u:y_c:y_t\sim 1:560:75000$,
one finds
\begin{align}
\epsilon_{d_R^1}:\epsilon_{d_R^2}:\epsilon_{d_R^3}&\sim 1:4:6.5
&\epsilon_{u_R^1}:\epsilon_{u_R^2}:\epsilon_{u_R^3}&\sim 1:110:600\,,
\label{RHhierarchy}
\end{align}
which shows that the down quark hierarchy is almost completely saturated  by the left handed suppression factors, $\epsilon_{q_L^i}$. In contrast,  the up quark sector  typically requires also a large RH hierarchy.

The most common flavor changing effects result from the coupling to KK gauge bosons (in particular, gluons). The wave functions of these modes rapidly approach a constant in the UV.
A flavor protection mechanism then arises naturally as follows.\cite{Huber:2003tu,Agashe:2004cp} The wave function overlap integral determines the interaction strength between the fermion current and the vector resonance schematically as 
\be
\mathcal L= g_{5d}\left( \frac{c_{(n)}}{\left|\log{\epsilon}\right|}+c'_{(n)}\epsilon_f^2\right) J_f^\mu A^{(n)}_\mu
\ee
where $c_{(n)}$ and $c_{(n)}'$ are $\mathcal O(1)$ numbers. 
Hence, for near UV localized fermions the flavor-nonuniversal second term is suppressed.
When rotating the fermions to the  mass eigenbasis, only this second term contributes to flavor changing couplings, while the first term remains flavor diagonal due to the unitarity of the rotations.
This way of suppressing flavor violation is sometimes referred to as the RS-GIM mechanism.
As we will see, although the RS-GIM mechanism greatly lowers the naive KK mass needed to suppress FCNCs, it is still around ~10--20 TeV (depending on the amount of fine-tuning accepted). Some authors therefore suggested additional flavor protection mechanisms, including symmetries,\cite{Cacciapaglia:2007fw,Cheung:2007bu,Santiago:2008vq,Csaki:2008eh,Bauer:2011ah,vonGersdorff:2012tt} or modified geometries,\cite{Atkins:2010cc,Archer:2011bk,Cabrer:2011qb,Archer:2012qa} which we will briefly comment on in the Sec.~\ref{boundsquarks} and \ref{boundsleptons}.

As is clear from this discussion, flavor violating couplings scale with the suppression factors $\epsilon_f$. Keeping the 4d Yukawas fixed, this means that one can achieve additional flavor protection by increasing the 5d Yukawa couplings $\hat Y_{ij}$. 
Typical scans allow $\hat Y_{ij}<3$, but larger values have been considered and allow for a significant reduction in the bounds. Let us thus briefly comment on naive dimensional analysis estimates for Yukawa couplings. Imposing   one loop corrections to not exceed tree level couplings, one would demand
\be
\hat Y^2 \hat \Lambda^{-2 d_Y}<\ell_d\,,
\ee
 where $\hat \Lambda$ is the cutoff in units of the warped-down curvature $\tilde k$. For instance, in order to be able to go up to  2 (3) KK gluon modes, one would impose $\hat\Lambda=5.5\ (8.6)$. 
The dimension $d_Y$ of the 5d Yukawa coupling depends on the nature of the Higgs (bulk or brane) and the location of the Yukawa coupling (for brane Higgs field $d_Y=-1$, while for a bulk Higgs field $d_Y=-\frac{1}{2}$ for a bulk coupling and $d_Y=-\frac{3}{2}$ for a brane coupling).   According to the location of the operator we need to apply a loop factor $\ell_5=24 \pi^3$ for bulk operators and $\ell_4=16 \pi^2$ for brane operators.\cite{vonGersdorff:2008df}
For a cutoff corresponding to 2 (3) KK modes, one then obtains $\hat Y<2.3\, (1.5)$ for a brane localized Higgs, while for a bulk Higgs one has $\hat Y_{\rm bulk}<11.6\, (9.3)$ and $\hat Y_{\rm brane}<1.0\, (0.5)$ for bulk and brane Yukawas respectively.
However, $\hat Y_{\rm bulk}$ also renormalizes $\hat Y_{\rm brane}$, which leads to a stronger bound on $\hat Y_{\rm bulk}$.
Imposing the bulk corrections to the brane coupling not to exceed the bound on $\hat Y_{\rm brane}$ leads to $\hat Y_{\rm bulk}^3<1.0\, (0.5)\,\ell_4$ or $\hat Y_{\rm bulk}<5.4\, (2.7)$. 
\footnote{One might object that this 5d reasoning is not fully self-consistent when one only considers very few KK modes, as 5d locality is not probed unless a sufficiently large number of KK modes is included. However we stick to this naive approach here to get a rough estimate of our perturbativity limits.} 
Related to these perturbativity bounds, it has also been pointed out that loop corrections to some flavor observables can become important as the latter are not suppressed but rather enhanced for larger Yukawa couplings.\cite{Csaki:2008zd}

Before starting the flavor analysis let us also mention recent re-evaluations of the electroweak precision observables. The bounds from $S$ and $T$ have become considerably stronger, here we quote the values found in Ref.~\refcite{fichet} from 
 the latest fit to EW data.\cite{Baak:2012kk} For the non-custodial case one finds from the S and T parameters
\be
m_{\mathbf{\rm KK}}|_{a=\infty}> 14.6~\textrm{TeV}\,,\quad m_{\mathbf{\rm KK}}|_{a=2}>8.0~\textrm{TeV}\,.
\label{noncust}
\ee
for brane and bulk Higgs respectively,
while in the custodial case, 
\be
m_{\mathbf{\rm KK}}|_{a=\infty}>7.6~\textrm{TeV}\,,\quad m_{\mathbf{\rm KK}}|_{a=2}>6.6~\textrm{TeV}\,.
\ee
In case of a bulk Higgs with $a=2$, one could argue that there is therefore no longer a strong motivation to introduce custodial symmetry, as the improvement in the bounds is only marginal. A further constraint arises from the $Zbb$ coupling; in the non-custodial case this requires roughly $c_{b_L}>0.4\ (0.45)$ for $\hat Y^b=1.0\ (3.0)$, in order to push the corresponding bound below $\sim 6$ TeV.\cite{Cabrer:2011qb} In the custodial case, there exists the possibility to relax the bound by imposing a discrete left-right parity $P_{LR}$.\cite{Agashe:2006at}

Moreover, from recent fits\cite{Dumont:2013wma} one also finds that the coupling $a_V$ of the Higgs to $W$ and $Z$ also severely constrains the KK scale. For instance, in the custodial case\cite{fichet}
\be
m_{\mathbf{\rm KK}}|_{a=\infty}>5.8~\textrm{TeV} \,\quad m_{\mathbf{\rm KK}}|_{a=2}>3.4~\textrm{TeV}
\ee

Finally, we point out that bounds from EWPT can be relaxed by reducing the coupling of the Higgs to the electroweak KK modes, for instance by adding brane localized kinetic terms\cite{Carena:2002dz} or by modifying the geometry in the IR.\cite{Falkowski:2008fz,Cabrer:2010si,Cabrer:2011fb} 
Notice that the $S$ parameter scales linearly with this coupling, while the $T$ parameter and $a_V$ scale quadratically.

In the remainder of the paper we will review bounds from several flavor observables, both for the custodial and non-custodial RS model. 
As a general caveat we stress that, contrary to bounds from electroweak precision data, which are insensitive to many of the free parameters of the models, flavor observables depend strongly on the $\mathcal O(1)$ 5d Yukawa couplings. Hence, typically, a scan over these parameters is appropriate, resulting in rather broad distributions of the allowed KK scales.\cite{Bauer:2009cf,Cabrer:2011qb} Where available, we quote percentiles of these distributions.
To get an order of magnitude estimate of the bounds, sometimes the expressions for the observables are estimated by an average 5d Yuakwa coupling $Y_*$. The advantage are simple expressions for the bounds, but they do not allow to reveal other important  quantitative features of the distributions such as variances and correlations between different observables.

\section{Operator analysis of flavor violation in RS}
\label{operators}

In this section we classify flavor violating operators in the unbroken electroweak basis. Short-distance contributions to flavor observables are conveniently accounted for in an effective field theory (EFT) approach with higher dimensional operators. As KK resonances are expected to be clearly separated from the electroweak (EW) scale, it is also convenient to compute these operators in the unbroken EW basis, assuming that the Higgs is a SM-like doublet. 
All relevant operators have dimension six and are hence suppressed by two inverse powers of the KK scale. We will not consider any operators of dimension larger than 6.
 
\subsection{Four fermion operators}
\label{4f}
Four fermion operators can be generated via the exchange of all spin-1 resonances such as KK gluons and electroweak KK modes. Moreover, in custodially protected models there is also the exchange of resonances related to the extended EW gauge group. They can contribute to both $\Delta F=2$ and $\Delta F=1$ processes. In models with a bulk Higgs boson they are also generated from KK Higgs exchange.
We will refer to these flavor violating operators as {\em proper four fermion vertices} (as opposed to those that are generated 
from $W$ and $Z$ exchange after EW breaking).

\subsection{Operators involving the Higgs}
\label{withhiggs}

There are three types of operators that can be generated at the tree level. The first class is given by
\be
i(\bar f_L^i\sigma^a\gamma^\mu f_L^j)(\phi^\dagger \sigma^a\overleftrightarrow D_\mu\phi)
\,, \qquad 
i(\bar f_{L,R}^i\gamma^\mu f_{L,R}^j)(\phi^\dagger \overleftrightarrow D_\mu\phi)
\ee
These operators are generated from KK exchange of electroweak gauge bosons as well as various fermionic KK resonances. After EW symmetry breaking they give rise to flavor changing couplings of the $Z$ and $W$ bosons. Notice that the $W$ boson coupling, as a result of the integration of the KK fermions, is no longer given by a unitary matrix.
We will refer to these corrections as {\em EW vertex corrections}.
They are typically not important for $\Delta F=2$ processes, as their contributions to the latter have an additional $m_{\rm KK}^2$ suppression with respect to the proper four-quark interactions described in Sec.~\ref{4f}.
On the other hand, for $\Delta F=1$ processes, the electroweak vertex corrections induce four fermion vertices (below the EW scale) which are enhanced by a factor of $\log\epsilon^{-1}\approx 37$ compared to the proper four fermion operators. The reason is that 
 the flavor preserving femionic coupling to KK modes in the former is volume suppressed compared to the Higgs coupling with the KK modes in the latter.

We also observe that the KK fermion contributions to the EW vertex corrections scale in a different way with the 5d Yukawa couplings compared to the KK-gauge contributions. Keeping the fermion masses fixed, the latter behave as $\sim Y_*^{-1}$, while the former scale linearly $\sim Y_*$.
We will comment on some implications of this fact in Sec.~\ref{boundsleptons}.
General expressions for the coefficients for the above operators from KK fermions have been given in Refs.~\refcite{Cabrer:2011qb,Carmona:2011rd}, see also Ref.~\refcite{Buras:2009ka}.
\footnote{Some authors\cite{Casagrande:2008hr,Bauer:2009cf} prefer not to integrate out KK fermions but instead deal with the infinite dimensional rotation matrices that result from diagonalizing the  mass matrix after EW breaking,  i.e.~solving the equations of motion in the broken phase. Both procedures must agree up to corrections $\sim\mathcal O(v^4/m_{\rm KK}^4)$. }

A second class is given by the operator
\be
i(\bar u_R^i\gamma^\mu d_R^j)(\tilde\phi^\dagger D_\mu\phi)
\label{anomalousW}
\ee
which gives rise to anomalous couplings of the RH quarks to the $W$ boson. These operators are generated by exchange of KK fermions as well as resonances of the extended EW gauge group in custodial models. They are suppressed for the light quarks due to their UV localization, but can still be important for flavor observables.

The third class is given by
\be
|\phi|^2\tilde\phi\,  \bar q_L^i u_R^j\,,\qquad |\phi|^2 \phi\,  \bar q_L^i d_R^j\,,\qquad 
|\phi|^2 \phi\,  \bar \ell_L^i e_R^j\,,
\ee
and are generated by KK fermion exchange. They lead to flavor violating couplings to the Higgs boson.

\subsection{Dipole operators}

Electroweak dipole operators
\be
\begin{array}{ccc}
\tilde \phi\, \bar q^i_L \sigma^{\mu\nu}u^j_R B_{\mu\nu}\,, 
&\phi\, \bar q^i_L \sigma^{\mu\nu}d^j_R B_{\mu\nu}\,, 
&\phi\, \bar \ell^i_L \sigma^{\mu\nu}e^j_R B_{\mu\nu}\,,\\
\vspace{-.25cm}
\\
\tilde\phi\, \sigma^a\bar q^i_L \sigma^{\mu\nu}u^j_R W^a_{\mu\nu}\,,\ 
&\ \phi\, \sigma^a\bar q^i_L \sigma^{\mu\nu}d^j_R W^a_{\mu\nu}\,,\ 
&\ \phi\, \sigma^a\bar \ell^i_L \sigma^{\mu\nu}e^j_R W^a_{\mu\nu}
\end{array}
\ee
and QCD dipole operators
\be
\tilde\phi\, \bar q^i_L t^a\,\sigma^{\mu\nu}u^j_R G^a_{\mu\nu}\,,\qquad
\phi\, \bar q^i_L t^a\,\sigma^{\mu\nu}d^j_R G^a_{\mu\nu}\,,
\ee
are only generated in loop diagrams and were computed e.g.~in Refs.~\refcite{Blanke:2012tv,Csaki:2010aj}.

\section{Bounds from the quark sector}
\label{boundsquarks}

\subsection{Bounds from CP violating $\Delta F=2$ observables}
\label{quarks}

By far the most stringent bounds arise from mixing in the neutral Kaon sector, in particular from $\epsilon_K$.~\cite{Huber:2003tu,Agashe:2004cp,Moreau:2006np,Csaki:2008zd,Santiago:2008vq,Csaki:2008eh,Blanke:2008zb,Agashe:2008uz,Bauer:2009cf,Cabrer:2011qb}
In RS models, there are contributions from  KK gluons to the Wilson coefficients $C^{sd}_i$ of the weak Hamiltonian. The largest impact on $
\epsilon_K$ then comes from $C_4^{sd}$, see for instance Ref.~\refcite{Bona:2007vi}. 
In the non-custodial model, distributions for the allowed values of KK gluon masses were computed in Refs~\refcite{Cabrer:2011qb,Bauer:2009cf}.  The 10, 20 and 50 percentiles are given by\cite{Cabrer:2011qb} 
\begin{align}
m_{\rm  KK}^{10\%}&=6.5{\rm\ TeV}&m_{\rm KK}^{20\%}&=9.7{\rm\ TeV}&
m_{\rm KK}^{50\%}&=19{\rm\ TeV}
\end{align}
The bounds apply to a maximally delocalized bulk Higgs field ($a=2$) and the scan used flat priors for the 5d Yukawa couplings with $|\hat Y_{ij}|<4$. 
A KK gluon of 3 TeV is compatible with the bounds from $\epsilon_K$  in roughly 2.5\% of the points of the scan, indicating that a fine-tuning of a few percent is needed in such models.
Since the EW sector KK resonances do not contribute to $C^{sd}_4$ and only give subleading contributions to $C^{sd}_1$, $\tilde C^{sd}_1$ and $C^{sd}_5$, the above bounds apply equally well to custodial models and hence can be considered fairly model independent. 
One should note however that in non-custodial models one needs to suppress too large deviations in the $Zbb$ coupling. Moreover, the latter constraints prefer actually small $\hat Y^d$, as there are contributions coming from the KK modes of the RH bottom quark.

Comparing the above limits with other references, one obtains roughly the same picture. For instance, Ref.~\refcite{Bauer:2009cf} analyses the case of a brane Higgs with $\hat Y<3$ and obtains the 10, 20 and 30 percentiles as \footnote{For better comparison we have translated here the definition $m_{\rm KK}\equiv \tilde k $ of   Ref.~\refcite{Bauer:2009cf}   into ours $m_{\rm KK}\equiv2.4 \tilde k$.}
\begin{align}
m_{\rm  KK}^{10\%}&=8.6{\rm\ TeV}&m_{\rm KK}^{20\%}&=12.5{\rm\ TeV}&
m_{\rm KK}^{30\%}&=20{\rm\ TeV}
\end{align}

We will see that other constraints in the quark sector are subleading, and even without fine tuning roughly consistent with KK gluons in the LHC range.
Given that the $\epsilon_K$ bounds are so severe, several approaches have been suggested in order to alleviate this fine tuning. We summarize them here briefly.
\begin{itemize}
\item {\em Alignment.\cite{Santiago:2008vq}} Since the $c_{d_R^i}$ of the down sector are so degenerate it has been suggested to make them exactly equal by imposing a $SU(3)_d$ symmetry, which can considerably lower the bounds.
\item {\em IR modifications of the metric.\cite{Archer:2011bk,Cabrer:2011qb} } 
IR modifications of the metric can alleviate the bounds by reducing flavour violating couplings between SM fields and KK states.
\item {\em Pseudo-axial gluons.\cite{Bauer:2011ah}} Adding an axial $SU(3)$ symmetry can help to exactly cancel the leading contributions to the Wilson coefficients from KK gluons.
\item {\em Modified matching of the strong coupling.} UV brane localized, negative  gluon kinetic terms allow to reduce the $SU(3)$ bulk gauge coupling and hence mildly lower the bounds.~\cite{Agashe:2008uz}
\end{itemize}

Further constraints arise in the up quark sector, in particular from CP violation in $D\bar D$ mixing.\cite{Gedalia:2009kh,Bauer:2009cf}
A rough estimate yields bounds on the KK gluon mass ranging from $m_{\rm K}=2.5 - 10$ TeV,\cite{Gedalia:2009kh} depending on the localization of the top quark; however, a quantitative estimate of the required fine-tuning as in the case of the $K\bar K$ mixing is not available. It is safe to say though that with current data these are the second most constraining quark flavor observables in the anarchic RS model.  
Let us stress that the RH up quark hierarchy Eq.~(\ref{RHhierarchy}) does not permit a simple alignment solution as in the case of the down sector.

Finally, CP violation in B mesons is subleading.\cite{Bauer:2009cf}
Let us close this section by remarking that the phenomenology in custodially symmetric models is somewhat different for $B$ and $D$ mesons, as electroweak corrections can become comparable with KK gluon contributions.\cite{Blanke:2008zb,Bauer:2009cf} 

\subsection{Bounds from CP conserving $\Delta F=2$ observables}

Analogous constraints from CP conserving quantities are much weaker. Bounds from the the $\Delta m_K$ $\Delta m_D$, $\Delta m_{B_d}$ and $\Delta m_{B_s}$ observables have been computed in e.g.~Ref.~\refcite{Csaki:2008zd} for the case of a brane Higgs field, where it was found that they do not lead to any significant tuning for a 3 TeV KK gluon. 
In contrast to the $K$ system, custodial KK modes can compete with KK gluons for the $B_{s,d}$ system.\cite{Blanke:2008zb}
See Refs.~\refcite{Casagrande:2008hr,Blanke:2008zb} for a more detailed discussion.

\subsection{Bounds from $\Delta F=1$ observables.}

In this section we give a brief summary of bounds resulting from $\Delta_F=1$ transitions. The bounds are in general much weaker, however, experimental sensitivity is expected to improve for many of these measurements in the forthcoming years, see
Ref.~\refcite{Agashe:2013kxa} for a recent survey of future sensitivity.

\begin{itemize}
\item
{\em Rare $K$ decays.}
Rare decays in the Kaon sector have for instance been considered in Refs.~\refcite{Blanke:2008yr} for the custodially symmetric model with brane localized Higgs field and in Ref.~\refcite{Straub:2013zca} in a two site model approximation. Fixing $m_{\rm KK}=2.45$ TeV and $\hat Y\leq 3$ and only considering points which satisfy the $\epsilon_K$ constraints it was found that the branching fraction $\mathcal B(K^0_L\to\pi_0\nu\bar\nu)$ can be enhanced by up to a factor of 5 compared to the SM, with the dominant contribution coming from the EW vertex corrections.\cite{Blanke:2008yr} Experimental sensitivity is however currently a factor of $10^3$ above the SM prediction. The experimental situation is a bit better for the $K_L^0\to \pi^0\ell^+\ell^-$ mode but still an order of magnitude away from the SM value, with little enhancement in the RS case.
The decay  $K^+\to\pi^+\nu\bar\nu$ can be enhanced by up to a factor of 2,\cite{Blanke:2008yr} which however is consistent with the current experimental error of $\mathcal O(100\%)$.
\item
{\em Rare $B$ decays.}
Rare decays for $B_s$ and $B_d$ mesons are typically small in models with $P_{LR}$ parity often imposed in custodial models in order to relax the bounds on $Z\to b\bar b$ decays. The same meachnism that suppresses the LH $Zb_L b_L$ coupling also efficiently suppresses the $Zd_L^i d_L^j$ couplings. In models without $P_{LR}$ symmetry some of the branching fractions, in particular $\mathcal B(B_s\to \mu^+\mu^-)$ can be enhanced,\cite{Blanke:2008yr} though no significant amount of fine tuning is needed for KK scales that ensure consistency with EWPT.\cite{Agashe:2013kxa}
\item
{\em Flavor violation in right handed $W$ couplings.} A model independent analysis constrains the coefficient $C^{tb}_{RR}$ of the operator Eqn.~(\ref{anomalousW}) as
\be
-0.0014<v^2 C^{tb}_{RR}<0.005
\ee
from the $b\to s\gamma$ branching ratio at 95\% C.L.\cite{Grzadkowski:2008mf} The RS prediction from integrating out KK modes of the left handed quarks is however suppressed as $C_{RR}\sim\epsilon_{b_R}\epsilon_{t_R} m_{\rm KK}^{-2}\sim 10^{-3}m_{\rm KK}^{-2}$ and hence the bounds are rather weak.\cite{Casagrande:2008hr} Exotic light fermion partners in custodial models are expected to only couple to either down or up sector and hence do not contribute to Eqn.~(\ref{anomalousW}).
\item
{\em Rare top decays.} $t\to c Z$ have been analyzed in Ref.~\refcite{Casagrande:2008hr} in the non-custodial model with brane localized Higgs. Their scan produces branching ratios ranging from $\mathcal B(t\to c Z)\approx 10^{-7} - 10^{-4}$ at $\tilde k=1.5$ TeV. CMS give a 95\% C.L. upper bound\cite{CMSTOP} $\mathcal B(t\to q Z)<7\times 10^{-4}$. Hence no significant fine tuning is necessary to ensure the experimental bounds, even for low KK scales.
\end{itemize}

\subsection{Bounds from dipole operators}

Bounds from loop induced dipole operators have for instance been considered in Refs.~\refcite{Agashe:2004cp,Moreau:2006np,Agashe:2008uz,Blanke:2012tv}. The very well measured $b\to s\gamma$ branching ratio\cite{Amhis:2012bh} 
\be
\mathcal B(B\to X_s\gamma)=(355\pm 24\pm 9)\times 10^{-6}
\label{bsg}
\ee
allows to put bounds on the KK scale. These operators are generated in penguin diagrams with various KK modes in the loop. A rough estimate yields\cite{Agashe:2008uz}
\be
m_{\rm KK}>0.63\,Y_* {\rm\ TeV}
\ee
where $m_{\rm KK}$ here refers to the Kaluza Klein fermions and $Y_*$ stands for the typical average Yukawa coupling, i.e. 
$Y_*=\langle \hat Y^d_{ij}\rangle$. 

Recently a more detailed analysis has been presented,~\cite{Blanke:2012tv} where also the effect of QCD dipole operators (that mix under RG flow with the electroweak ones) are included. For a IR scale $\tilde k=1$ TeV (corresponding to 2.5 TeV KK gauge bosons) as well as $\hat Y<3$, no substantial portion of the RS parameter space lies outside the experimentally allowed region, Eqn.~(\ref{bsg}), both for the custodial and non-custodial models. For the custodial model, only $\sim$15\% of the parameter space were found to be excluded. One would expect this portion to further reduce significantly for KK scales consistent with EWPT.

\section{Bounds from the lepton sector}
\label{boundsleptons}

The lepton sector has been considered e.g.~in Refs.~\refcite{Huber:2003tu,Moreau:2006np,Agashe:2006iy,Csaki:2010aj,Iyer:2012db}.

The most stringent bounds in the lepton sector arise from the $\mu\to e\gamma$ decay mode. 
Taking into account the most recent MEG bound \cite{Adam:2013mnn}
\be
\mathcal B(\mu\to e\gamma) < 5.7\times 10^{-13}\ (90 \%\ {\rm C.L.})
\ee
we have rescaled results\cite{Agashe:2006iy,Csaki:2010aj} based on older data accordingly.
Ref.~\refcite{Agashe:2006iy} analyzes the  the non-custodial model with bulk Higgs ($a=2$)
and quotes the rough estimate
\be
m_{\rm KK}>17.1\ (33.8)\, {\rm TeV}
\ee
for the two values $Y_*=1$ ($Y_*=2$) respectively.

The case of brane-localized Higgs field has been analyzed in Ref.~\refcite{Csaki:2010aj}, both for the custodial and non-custodial models. It is found that 
\be
m_{\rm KK}>|\alpha Y_*^2+\beta|^\frac{1}{2} 52\, {\rm TeV}
\ee
where $ \alpha=-0.065$ ($-0.15$)  in the noncustodial (custodial) model and $\beta\lesssim 0.03$.

Furthermore, there are bounds from the decay $\mu\to 3e$ and from $\mu\to e$ conversion in nuclei. At energies below the EW scale, they can be accounted for by four-lepton or two-lepton two-quark operators respectively. 
Being $\Delta F=1$ transitions, they are dominated by electroweak vertex corrections, as discussed in Sec.~\ref{withhiggs}, in other words they are due to flavor changing $Z$ couplings coming from the operators
\be
\mathcal O_{L}=i(\bar \mu_L\sigma^3\gamma^\mu e_L)(\phi^\dagger \sigma^3\overleftrightarrow D_\mu\phi)
\,, \qquad 
\mathcal O'_{\chi}=i(\mu_{\chi}\gamma^\mu e_{\chi})(\phi^\dagger \overleftrightarrow D_\mu\phi)
\ee
In the model with IR brane localized Higgs field, the bounds from KK exchange of the EW gauge sector can be roughly estimated as\cite{Csaki:2010aj}
\be
m_{\rm KK}>6.0\,Y_*^{-\frac{1}{2}}\,{\rm TeV}\,,\qquad m_{\rm KK}>3.4\,Y_*^{-\frac{1}{2}}\,{\rm TeV}
\label{trilepton}
\ee
for the non-custodial and custodial models respectively. 

The bounds from LFV observables are summarized in Table \ref{LFV}. Let us reiterate that the distribution of bounds are rather broad and the quoted numbers should be taken as indicative only. As a matter of fact, examining by eye the scans of Ref.~\refcite{Agashe:2006iy}, the typical bounds seem to be stronger than the naive estimates. Bounds on $\tau$ decays are much weaker and only give subleading bounds.\cite{Agashe:2006iy}
\begin{table}[h]
\tbl{ Summary of estimates for bounds from LFV. The $\mu\to e\gamma$ constraints have been updated according to the latest constraints from 
MEG. The two values apply to $Y_*=1\ (Y_*=2)$. We have set the parameter $\beta$ to zero.
\label{LFV}
}
{\begin{tabular}{@{}cccc@{}} \toprule
Process & Min., Bulk Higgs\cite{Agashe:2006iy} & Min., Brane Higgs\cite{Csaki:2010aj} & Cust., Brane Higgs\cite{Csaki:2010aj}  \\
\colrule
$\mu\to e$	&6.7 (4.7) TeV	& 6 (4.2) TeV & 3.4 (2.4) TeV\\
$\mu\to e\gamma$	& 17.1 (33.8) TeV& 13.4 (26.8) TeV & 20.3 (40.6) TeV
\\ 
\botrule
\end{tabular}\label{ta1} }
\end{table}

Several authors have proposed models based on discrete\cite{Csaki:2008qq,delAguila:2010vg,Kadosh:2010rm,Hagedorn:2011un}
 or continuous\cite{Perez:2008ee,vonGersdorff:2012tt} symmetries that can significantly reduce the bounds from $\mu\to e\gamma$ (see also Refs.~\refcite{Atkins:2010cc,Agashe:2009tu} for some alternative proposals).
It should also be noted that the constraints from the $\mu\to e \gamma$ rate is somewhat more model-dependent than the quark constraints, as it depends to some extend on the nature of neutrino masses.

Let us close this section with an observation related to the tree level mediated processes. The conventional wisdom is that the trilepton decay and $\mu \to e$ conversion rates scale inversely with the 5d Yukawa couplings once the physical 4d Yukawa couplings are held fixed.\cite{Agashe:2006iy,Csaki:2010aj}
\footnote{This is in contrast to the loop induced $\mu\to e\gamma$ rate, which grows with the 5d Yukawa coupling. It has been noted that therefore there exists a tension between the tree level and loop level observables.\cite{Agashe:2006iy,Csaki:2010aj}} 
However, we also expect the exchange of the KK fermions to be important for larger values of the Yukawa couplings, as discussed in Sec.~\ref{withhiggs}, a contribution that grows linearly with the 5d Yukawa couplings. We are not aware of any analysis that takes into account this contribution. 
Using the expressions of Ref.~\refcite{Cabrer:2011qb}, 
and working in the non-custodial model for definiteness,
we find the KK-fermion contribution to exceed the KK-gauge contribution for $Y_*\approx 3$ ($Y_*\approx 6$) for $a=\infty$ ($a=2$) respectively, for both the left and right handed $Z\mu e$ couplings, which, although mostly subleading, could give some effect and should be included in numerical scans. Moreover, custodial models typically contain light vector-like partners of the $\tau$ \cite{delAguila:2010es} whose exchange also contributes to the $Z\mu e$ couplings and which should give much stronger effects that could easily overwhelm the KK gauge contributions, even for moderate Yukawa couplings.

\section{Conclusions}

Suppression of flavor violation in WED is very efficient for most observables in both quark and lepton sectors. 
Fixing the KK masses to $\sim 5-6$ TeV, as required by EWPT in the most favourable cases, most of the anarchic parameter space of WED (i.e., the $\mathcal O(1)$ 5d Yukawa couplings and bulk mass parameters $c_f$ that fix the fermion masses and mixings) is compatible with almost all flavor violating observables, with the exception of $\epsilon_K$ in the quark sector, as well as $\mu\to e \gamma$ in the lepton sector. The constraint from the former can be satisfied at the cost of a fine tuning of a few percent, while those of the latter are more severe. This clearly points towards a non-minimal realization of the lepton flavor sector, possibly requiring the existence of either discrete or continuous symmetries. 

\appendix

\section*{Acknowledgments}
I would like to thank the Funda\c{c}\~ao de
Amparo \`a Pesquisa do Estado de S\~ao Paulo (FAPESP) for financial support.



\begin{thebibliography}{0}

\bibitem{Randall:1999ee}
  L.~Randall and R.~Sundrum,
  Phys.\ Rev.\ Lett.\  {\bf 83} (1999) 3370
  [hep-ph/9905221].

\bibitem{Grossman:1999ra}
  Y.~Grossman and M.~Neubert,
  Phys.\ Lett.\ B {\bf 474} (2000) 361
  [hep-ph/9912408].

\bibitem{Gherghetta:2000qt}
  T.~Gherghetta and A.~Pomarol,
  Nucl.\ Phys.\ B {\bf 586} (2000) 141
  [hep-ph/0003129].
  
\bibitem{Huber:2000ie}
  S.~J.~Huber and Q.~Shafi,
  Phys.\ Lett.\ B {\bf 498} (2001) 256
  [hep-ph/0010195].

\bibitem{Cabrer:2011qb}
  J.~A.~Cabrer, G.~von Gersdorff and M.~Quiros,
  JHEP {\bf 1201} (2012) 033
  [arXiv:1110.3324 [hep-ph]].

\bibitem{Huber:2003tu}
  S.~J.~Huber,
  Nucl.\ Phys.\ B {\bf 666} (2003) 269
  [hep-ph/0303183].
  
\bibitem{Agashe:2004cp}
  K.~Agashe, G.~Perez and A.~Soni,
  Phys.\ Rev.\ D {\bf 71} (2005) 016002
  [hep-ph/0408134].

\bibitem{Cacciapaglia:2007fw}
  G.~Cacciapaglia, C.~Csaki, J.~Galloway, G.~Marandella, J.~Terning and A.~Weiler,
  JHEP {\bf 0804} (2008) 006
  [arXiv:0709.1714 [hep-ph]].

\bibitem{Cheung:2007bu}
  C.~Cheung, A.~L.~Fitzpatrick and L.~Randall,
  JHEP {\bf 0801} (2008) 069
  [arXiv:0711.4421 [hep-th]].


\bibitem{Santiago:2008vq}
  J.~Santiago,
  JHEP {\bf 0812} (2008) 046
  [arXiv:0806.1230 [hep-ph]].
  
\bibitem{Csaki:2008eh}
  C.~Csaki, A.~Falkowski and A.~Weiler,
  Phys.\ Rev.\ D {\bf 80} (2009) 016001
  [arXiv:0806.3757 [hep-ph]].

\bibitem{Bauer:2011ah}
  M.~Bauer, R.~Malm and M.~Neubert,
  Phys.\ Rev.\ Lett.\  {\bf 108} (2012) 081603
  [arXiv:1110.0471 [hep-ph]].



\bibitem{vonGersdorff:2012tt}
  G.~von Gersdorff, M.~Quiros and M.~Wiechers,
  JHEP {\bf 1302} (2013) 079
  [arXiv:1208.4300 [hep-ph]].




\bibitem{Atkins:2010cc}
  M.~Atkins and S.~J.~Huber,
  Phys.\ Rev.\ D {\bf 82} (2010) 056007
  [arXiv:1002.5044 [hep-ph]].

\bibitem{Archer:2011bk}
  P.~R.~Archer, S.~J.~Huber and S.~Jager,
  JHEP {\bf 1112} (2011) 101
  [arXiv:1108.1433 [hep-ph]].
  
\bibitem{Archer:2012qa}
  P.~R.~Archer,
  JHEP {\bf 1209} (2012) 095
  [arXiv:1204.4730 [hep-ph]].

\bibitem{vonGersdorff:2008df}
  G.~von Gersdorff,
  JHEP {\bf 0808} (2008) 097
  [arXiv:0805.4542 [hep-th]].


\bibitem{Csaki:2008zd}
  C.~Csaki, A.~Falkowski and A.~Weiler,
  JHEP {\bf 0809} (2008) 008
  [arXiv:0804.1954 [hep-ph]].



\bibitem{fichet}
	S.~Fichet and G.~von Gersdorff, to appear
	
\bibitem{Baak:2012kk}
  M.~Baak, M.~Goebel, J.~Haller, A.~Hoecker, D.~Kennedy, R.~Kogler, K.~Moenig and M.~Schott {\it et al.},
  Eur.\ Phys.\ J.\ C {\bf 72} (2012) 2205
  [arXiv:1209.2716 [hep-ph]].
  
  
\bibitem{Agashe:2006at}
  K.~Agashe, R.~Contino, L.~Da Rold and A.~Pomarol,
  Phys.\ Lett.\ B {\bf 641} (2006) 62
  [hep-ph/0605341].
  
  
 
\bibitem{Dumont:2013wma} 
  B.~Dumont, S.~Fichet and G.~von Gersdorff,
  JHEP {\bf 1307}, 065 (2013)
  [arXiv:1304.3369 [hep-ph]].
  
  

\bibitem{Carena:2002dz}
  M.~S.~Carena, E.~Ponton, T.~M.~P.~Tait and C.~E.~MWagner,
  Phys.\ Rev.\ D {\bf 67} (2003) 096006
  [hep-ph/0212307].

  
  
\bibitem{Falkowski:2008fz}
  A.~Falkowski and M.~Perez-Victoria,
  JHEP {\bf 0812} (2008) 107
  [arXiv:0806.1737 [hep-ph]].
  
\bibitem{Cabrer:2010si}
  J.~A.~Cabrer, G.~von Gersdorff and M.~Quiros,
  Phys.\ Lett.\ B {\bf 697} (2011) 208
  [arXiv:1011.2205 [hep-ph]].
  
\bibitem{Cabrer:2011fb}
  J.~A.~Cabrer, G.~von Gersdorff and M.~Quiros,
  JHEP {\bf 1105} (2011) 083
  [arXiv:1103.1388 [hep-ph]].
      
  
  
\bibitem{Bauer:2009cf}
  M.~Bauer, S.~Casagrande, U.~Haisch and M.~Neubert,
  JHEP {\bf 1009} (2010) 017
  [arXiv:0912.1625 [hep-ph]].
  

\bibitem{Carmona:2011rd}
  A.~Carmona and J.~Santiago,
  JHEP {\bf 1201} (2012) 100
  [arXiv:1110.5651 [hep-ph]].
  

\bibitem{Buras:2009ka}
  A.~J.~Buras, B.~Duling and S.~Gori,
  JHEP {\bf 0909} (2009) 076
  [arXiv:0905.2318 [hep-ph]].
  
\bibitem{Casagrande:2008hr}
  S.~Casagrande, F.~Goertz, U.~Haisch, M.~Neubert and T.~Pfoh,
  JHEP {\bf 0810} (2008) 094
  [arXiv:0807.4937 [hep-ph]].

\bibitem{Blanke:2012tv}
  M.~Blanke, B.~Shakya, P.~Tanedo and Y.~Tsai,
  JHEP {\bf 1208} (2012) 038
  [arXiv:1203.6650 [hep-ph]].
  
  \bibitem{Csaki:2010aj}
  C.~Csaki, Y.~Grossman, P.~Tanedo and Y.~Tsai,
  Phys.\ Rev.\ D {\bf 83} (2011) 073002
  [arXiv:1004.2037 [hep-ph]].



  
  
  
  
  
  
\bibitem{Moreau:2006np}
  G.~Moreau and J.~I.~Silva-Marcos,
  JHEP {\bf 0603} (2006) 090
  [hep-ph/0602155].
  
  


  
  
  

\bibitem{Blanke:2008zb}
  M.~Blanke, A.~J.~Buras, B.~Duling, S.~Gori and A.~Weiler,
  JHEP {\bf 0903} (2009) 001
  [arXiv:0809.1073 [hep-ph]].

  
\bibitem{Agashe:2008uz}
  K.~Agashe, A.~Azatov and L.~Zhu,
  Phys.\ Rev.\ D {\bf 79} (2009) 056006
  [arXiv:0810.1016 [hep-ph]].


\bibitem{Bona:2007vi}
  M.~Bona {\it et al.}  [UTfit Collaboration],
  JHEP {\bf 0803} (2008) 049
  [arXiv:0707.0636 [hep-ph]].
  
  
  
\bibitem{Gedalia:2009kh}
  O.~Gedalia, Y.~Grossman, Y.~Nir and G.~Perez,
  Phys.\ Rev.\ D {\bf 80} (2009) 055024
  [arXiv:0906.1879 [hep-ph]].

\bibitem{Agashe:2013kxa}
  K.~Agashe, M.~Bauer, F.~Goertz, S.~J.~Lee, L.~Vecchi, L.~-T.~Wang and F.~Yu,
  arXiv:1310.1070 [hep-ph].

\bibitem{Blanke:2008yr}
  M.~Blanke, A.~J.~Buras, B.~Duling, K.~Gemmler and S.~Gori,
  JHEP {\bf 0903} (2009) 108
  [arXiv:0812.3803 [hep-ph]].
  
\bibitem{Straub:2013zca}
  D.~M.~Straub,
  JHEP {\bf 1308} (2013) 108
  [arXiv:1302.4651 [hep-ph]].

  
\bibitem{Grzadkowski:2008mf}
  B.~Grzadkowski and M.~Misiak,
  Phys.\ Rev.\ D {\bf 78} (2008) 077501
   [Erratum-ibid.\ D {\bf 84} (2011) 059903]
  [arXiv:0802.1413 [hep-ph]].


\bibitem{CMSTOP}
CMS PAS TOP-12-037

\bibitem{Amhis:2012bh}
  Y.~Amhis {\it et al.}  [Heavy Flavor Averaging Group Collaboration],
  arXiv:1207.1158 [hep-ex].

  \bibitem{Agashe:2006iy}
  K.~Agashe, A.~E.~Blechman and F.~Petriello,
  Phys.\ Rev.\ D {\bf 74} (2006) 053011
  [hep-ph/0606021].

\bibitem{Iyer:2012db}
  A.~MIyer and S.~KVempati,
  Phys.\ Rev.\ D {\bf 86} (2012) 056005
  [arXiv:1206.4383 [hep-ph]].

\bibitem{Adam:2013mnn}
  J.~Adam {\it et al.}  [MEG Collaboration],
  arXiv:1303.0754 [hep-ex].
   
\bibitem{Csaki:2008qq}
  C.~Csaki, C.~Delaunay, C.~Grojean and Y.~Grossman,
  JHEP {\bf 0810} (2008) 055
  [arXiv:0806.0356 [hep-ph]].
  
\bibitem{delAguila:2010vg}
  F.~del Aguila, A.~Carmona and J.~Santiago,
  JHEP {\bf 1008} (2010) 127
  [arXiv:1001.5151 [hep-ph]].
  
\bibitem{Kadosh:2010rm}
  A.~Kadosh and E.~Pallante,
  JHEP {\bf 1008} (2010) 115
  [arXiv:1004.0321 [hep-ph]].
  
\bibitem{Hagedorn:2011un}
  C.~Hagedorn and M.~Serone,
  JHEP {\bf 1110} (2011) 083
  [arXiv:1106.4021 [hep-ph]].

\bibitem{Perez:2008ee}
  G.~Perez and L.~Randall,
  JHEP {\bf 0901} (2009) 077
  [arXiv:0805.4652 [hep-ph]].
  
 

  
\bibitem{Agashe:2009tu}
  K.~Agashe,
  Phys.\ Rev.\ D {\bf 80} (2009) 115020
  [arXiv:0902.2400 [hep-ph]].


  
\bibitem{delAguila:2010es}
  F.~del Aguila, A.~Carmona and J.~Santiago,
  Phys.\ Lett.\ B {\bf 695} (2011) 449
  [arXiv:1007.4206 [hep-ph]].






  
  



  
  

  

  

  

  
  


\end{thebibliography}
\end{document}